\documentclass[twocolumn,prl,showpacs]{revtex4}
\usepackage{graphicx}
\usepackage{dcolumn}
\usepackage{bm}
\usepackage{amsmath}
\usepackage{times}

\begin{document}

\preprint{Odd-frequency proximity}
\title{Odd-frequency pairing and Josephson effect at superconducting interfaces}
\author{ Y. Tanaka$^{1,2}$  A. A. Golubov$^{3}$, S. Kashiwaya$^{4}$ and M. Ueda$^{5}$}
\affiliation{$^1$Department of Applied Physics, Nagoya University, Nagoya, 464-8603,
Japan \\
$^2$ CREST Japan Science and Technology Cooperation (JST) 464-8603 Japan \\
$^3$ Faculty of Science and Technology, University of Twente, The
Netherlands \\
$^4$ National Institute of Advanced Industrial Science and Technology
(AIST), Tsukuba, 305-8568, Japan \\
$^5$ Department of Physics, Tokyo Institute of Technology, Tokyo,
152-8551, Japan \\
}
\date{\today}

\begin{abstract}
We demonstrate that, quite generally, the spin-singlet even-parity 
(spin-triplet odd-parity) pair potential in a superconductor induces the
odd-frequency pairing component with spin-singlet odd-parity
(spin-triplet even-parity) near interfaces.
The magnitude of the induced odd-frequency
component is enhanced in the presence of the midgap
Andreev resonant state due to the sign change of the anisotropic
pair potential at the interface.
The Josephson effect should therefore occur
between odd- and even-frequency superconductors, 
contrary to the standard wisdom. 
A method to probe the odd-frequency superconductors is proposed.

\end{abstract}

\pacs{74.45.+c, 74.50.+r, 74.20.Rp}
\maketitle



%
%

%



It is well known that Josephson coupling occurs between superconductors with the same
order parameter symmetry \cite {Sigrist}. Generally, symmetries with respect to momentum, spin and
time are considered. On the basis of symmetry with respect to time, two classes of superconductors
are introduced, 
referred to as odd-frequency and even-frequency superconductors.
In accordance with the Pauli principle, the even-frequency state is characterized by spin-singlet
even-parity or spin-triplet odd-parity order parameter, while odd-frequency superconductors belong
to spin-singlet odd-parity or spin-triplet even-parity pairing state.
It was suggested in Ref. \cite{Abraham} from basic symmetry arguments that the first order
Josephson coupling should be absent between odd- and even-frequency superconductors.
However, as shown in the present paper, an odd-frequency component of the pair potential
is quite generally induced at interfaces in superconductors with even-frequency bulk
pair potential, while an even-frequency component is induced at interfaces
in odd-frequency superconductors. Therefore Josephson coupling between even-
and odd-frequency superconductors should be possible. \par
Up to now, almost all of known superconductors belong to the
symmetry class of the even-frequency pairing. The possibility of the 
odd-frequency pairing state in a uniform system was proposed for  
$^{3}$He in \cite{Berezinskii} and for a superconducting state
involving strong correlations \cite{Balatsky,Abraham,Vojta,Fuseya}.
%
%
Recently, the realization of the odd-frequency pairing state in
inhomogeneous systems was proposed by Ref. \onlinecite{Efetov1} in
ferromagnet/superconductor heterostructures. This issue was further
addressed in several related studies \cite{Efetov2,Kadigrob}. Recent
experiments provided evidence for such anomalous pairing states
\cite{expt}. Furthermore, two of the present authors predicted that
the odd-frequency pairing state can be induced in a diffusive normal
metal attached to a spin-triplet superconductor \cite{Golubov2006}.
However, in these examples the odd-frequency pairing is realized due
to spin-triplet correlations. A question naturally arises whether the
spin-triplet ordering is a necessary condition for the existence of
the odd-frequency pairing state at superconducting interfaces.
In order to address this issue, one should consider self-consistently the
pair potential near the interface.

The classic example of an inhomogeneous superconducting system is
a ballistic normal metal/superconductor (N/S) junction with an
even-frequency superconductor. Particularly interesting are the
cases when S is a spin-singlet $d$-wave or a spin-triplet $p$-wave
superconductor
%
%
since the sign change of the pair potential probed by quasiparticles
injected and reflected by the N/S interface can occur in such
junctions. As a result, the pair potential is suppressed near the
interface \cite{Spatial} and the midgap Andreev resonant states
(MARS) are formed \cite{TK95}.
%
The appearance of unusual charge transport in the presence of MARS
\cite{TK95} suggests the presence of underlying anomalous pairing
states.

In this Letter we address the issue of odd-frequency pairing
in the generic case of a N/S interface, where superconductor S has
an even-frequency pairing state in the bulk. We will use the
quasiclassical Green's function formalism where the spatial
dependence of the pair potential is determined self-consistently.
We will show that, quite generally, the odd-frequency component is
induced near superconducting interfaces due to the spatial variation
of the pair potential \cite{Eschrig2006}.
%
If a superconductor
has an ESE or ETO pairing state in the bulk, the order parameter at the
interface has respectively an odd-frequency spin-singlet odd-parity
(OSO) or an odd-frequency spin-triplet even-parity (OTE).
In the absence of MARS (e.g. ESE superconductor with $s$-wave or
$d_{x^{2}-y^{2}}$-wave symmetry of the order parameter), the
magnitude of the odd-frequency component of the pair amplitude
decreases with a decrease of the transmission coefficient $T_{m}$
across the interface.
On the other hand, in the presence of MARS (e.g. ETO superconductor
with $p_{x}$-wave or ESE one with $d_{xy}$-wave symmetry of the
order parameter), the magnitude of the odd-frequency component is
enhanced with a decrease of $T_{m}$.
Similarly, in the case of a bulk odd-frequency superconductor, an
even-frequency component of the order parameter is induced at the
interface.
An important application of the above results is the existence of Josephson coupling
between bulk odd- and even-frequency superconductors to the
first order in $T_{m}$.
 \par

In the following, we consider an N/S junction as the simplest example
of a non-uniform superconducting system without impurity scattering.
Both the ESE and ETO
states are considered in the superconductor. As regards the
spin-triplet superconductor, we choose $S_{z}=0$ for the simplicity.
We assume a thin insulating barrier located at the
N/S interface ($x=0$) with N ($x<0)$ and S($x>0$)
modeled by a delta function $H\delta (x)$, where $ H $ is
the magnitude of the strength of the delta function potential. The
reflection coefficient of the junction for the quasiparticle for
the injection angle $\theta $ is given by $R=Z^{2}
/(Z^{2}+4\cos ^{2}\theta )=1-T_{m}$ with $Z=2H/v_{F}$, where $\theta $
$(-\pi /2<\theta <\pi /2)$ is measured from the normal to the
interface and $v_{F}$ is the Fermi velocity. 
The quasiclassical
Green's function in a superconductor is parameterized 
in terms of 
\begin{equation}
\hat{g}_{\pm }=f_{1\pm }\hat{\tau}_{1}+f_{2\pm }\hat{\tau}_{2}+g_{\pm }\hat{%
\tau}_{3},\ \ \hat{g}_{\pm }^{2}=\hat{1}
\end{equation}%
with Pauli matrices $\hat{\tau}_{i}$ ($i=1-3$) and unit matrix $\hat{1}$. Here,
the index $+$ ($-$) denotes the left (right) going quasiparticles \cite%
{Serene}. It is possible to express the above Green's function as
$f_{1\pm }=\pm i(F_{\pm }+D_{\pm })/(1-D_{\pm }F_{\pm })$, $f_{2\pm
}=-(F_{\pm }-D_{\pm })/(1-D_{\pm }F_{\pm })$, and $g_{\pm
}=(1+D_{\pm }F_{\pm })/(1-D_{\pm }F_{\pm })$, where $D_{\pm }$ and
$F_{\pm }$ satisfy the Eilenberger equations in the Riccatti
parameterization \cite{Ricatti}
\begin{equation}
v_{Fx}\partial _{x}D_{\pm }=-\bar{\Delta}_{\pm }(x)(1-D_{\pm }^{2})+2\omega
_{n}D_{\pm }  \label{eq.1a}
\end{equation}%
\begin{equation}
v_{Fx}\partial _{x}F_{\pm }=-\bar{\Delta}_{\pm }(x)(1-F_{\pm }^{2})-2\omega
_{n}F_{\pm }.  \label{eq.1b}
\end{equation}%
where $v_{Fx}$ is the $x$ component of the Fermi velocity, $\omega _{n}=2\pi T(n+1/2)$ is
the Matsubara frequency, with temperature $T$.
$\bar{\Delta}_{+}(x)$ ($\bar{\Delta}_{-}(x)$)
is the effective pair potential for left going (right going) quasiparticles.
%
%
Since the interface is flat, $F_{\pm }=-RD_{\mp }$ holds at $x=0$
\cite{Ricatti}. Here we consider the situation without mixing of
different symmetry channels for the pair potential. Then
$\bar{\Delta}_{\pm }(x)$ is expressed by $\bar{\Delta}_{\pm
}(x)=\Delta (x)\Phi _{\pm }(\theta )\Theta(x)$ with the form factor $\Phi
_{\pm }(\theta )$ given by 1, $\cos 2\theta $, $\pm \sin 2\theta $,
$\pm \cos \theta $, and $\sin \theta $ for $s$, $d_{x^{2}-y^{2}}$ ,
$d_{xy}$, $p_{x}$ and $p_{y}$-wave superconductors, respectively.
$\Delta
(x)$ is determined by the self-consistency equation%
\begin{equation}
\Delta (x)=\frac{2 T}{\mathrm{log}\frac{T}{T_{C}}+\displaystyle%
\sum_{n\geq 1}\frac{1}{n-\frac{1}{2}}}\displaystyle\sum_{n\geq 0}\int_{-\pi
/2}^{\pi /2}d\theta G(\theta )f_{2+}
\end{equation}%
with $G(\theta )=1$ for the $s$-wave case and $G(\theta )=2\Phi (\theta )$ for
other cases, respectively \cite{Matsumoto}.
$T_{C}$ is the transition temperature of the superconductor.
%
The condition in the bulk is $\Delta (\infty )=\Delta $. Since the pair
potential $\bar{\Delta}(x)$ is a real number, the resulting $f_{1\pm }$ is
an imaginary and $f_{2\pm }$ is a real number.

In the following, we explicitly write $f_{1\pm }=f_{1\pm }(\omega
_{n},\theta )$, $f_{2\pm }=f_{2\pm }(\omega _{n},\theta )$, $F_{\pm }=F_{\pm
}(\omega _{n},\theta )$ and $D_{\pm }=D_{\pm }(\omega _{n},\theta )$.
For $%
x=\infty $,
we obtain $f_{1\pm }(\omega _{n},\theta )=0$ and $%
f_{2\pm }(\omega _{n},\theta )=\frac{\Delta \Phi _{\pm }(\theta )}{\sqrt{%
\omega _{n}^{2}+\Delta ^{2}\Phi _{\pm }^{2}(\theta _{\pm })}}$. Note that $%
f_{1\pm }(\omega _{n},\theta )$ becomes finite due to the spatial variation
of the pair potential and does not exist as the bulk. From eqs. (\ref{eq.1a}%
) and (\ref{eq.1b}), we can show $D_{\pm }(-\omega _{n},\theta )=1/D_{\pm
}(\omega _{n},\theta )$ and $F_{\pm }(-\omega _{n},\theta )=1/F_{\pm
}(\omega _{n},\theta )$. After a simple manipulation, we obtain $f_{1\pm
}(\omega _{n},\theta )=-f_{1\pm }(-\omega _{n},\theta )$ and $f_{2\pm
}(\omega _{n},\theta )=f_{2\pm }(-\omega _{n},\theta )$ for any $x$. It is
remarkable that functions $f_{1\pm }(\omega _{n},\theta )$ and $f_{2\pm
}(\omega _{n},\theta )$ correspond to an odd-frequency and an even-frequency
one of the pair amplitude, respectively \cite{Efetov2}.

Next, we shall discuss the parity of these pair amplitudes. The even-parity
(odd-parity) pair amplitude should satisfy the relation $f_{i\pm }(\omega
_{n},\theta )=f_{i\mp }(\omega _{n},-\theta )$ ($f_{i\pm }(\omega
_{n},\theta )=-f_{i\mp }(\omega _{n},-\theta )$), with $i=1,2$. For an
even(odd)-parity superconductor $\Phi _{\pm }(-\theta )=\Phi _{\mp }(\theta
) $ ($\Phi _{\pm }(-\theta )=-\Phi _{\mp }(\theta ))$. Then, we can show
that $D_{\pm }(-\theta )=D_{\mp }(\theta )$ and $F_{\pm }(-\theta )=F_{\mp
}(\theta )$ for  even-parity case and $D_{\pm }(-\theta )=-D_{\mp }(\theta )$
and $F_{\pm }(-\theta )=-F_{\mp }(\theta )$ for odd-parity case,
respectively. The resulting $f_{1\pm }(\omega _{n},\theta )$ and $f_{2\pm
}(\omega _{n},\theta )$ satisfy $f_{1\pm }(\omega _{n},\theta )=-f_{1\mp
}(\omega _{n},-\theta )$ and $f_{2\pm }(\omega _{n},\theta )=f_{2\mp
}(\omega _{n},-\theta )$ for an even-parity superconductor and $f_{1\pm
}(\omega _{n},\theta )=f_{1\mp }(\omega _{n},-\theta )$ and $f_{2\pm
}(\omega _{n},\theta )=-f_{2\mp }(\omega _{n},-\theta )$ for an odd-parity
superconductor. Note that the parity of the odd-frequency component $f_{1\pm
}(\omega _{n},\theta )$ is different from that of the bulk superconductor
for all cases.

Let us now focus on the values of the pair amplitudes at the interface $x=0$%
. We concentrate on two extreme cases with (1) $\Phi _{+}(\theta
)=\Phi _{-}(\theta )$ and (2) $\Phi _{+}(\theta )=-\Phi _{-}(\theta
)$. In the first case, MARS is absent since there is no sign change
of the pair potential felt by the quasiparticle at the interface.
Then $D_{+}=D_{-}$ is satisfied. On the other hand, in the second
case, MARS is generated near the interface due to the sign change of
the pair potential. Then $D_{+}=-D_{-}$ is satisfied \cite{TK95}. At
the interface, it is easy to show that $f_{1\pm }=\pm
i(1-R)D_{+}/(1+RD_{+}^{2})$ and $f_{2\pm
}=(1+R)D_{+}/(1+RD_{+}^{2})$ for Case (1) and $f_{1\pm
}=i(1+R)D_{+}/(1-RD_{+}^{2})$ and $f_{2\pm }=\pm
(1-R)D_{+}/(1-RD_{+}^{2})$ for Case (2), respectively, where the
real number $ D_{+}$ satisfies $\mid D_{+}\mid <1$ for $\omega
_{n}\neq 0$. For Case (1), the magnitude of $f_{1\pm }$ is always
smaller than that of $f_{2\pm }$. For Case (2),
the situation is reversed. In the low transparent limit with
$R\rightarrow 0$, only the $f_{1\pm }$ is nonzero. Namely, only the
even-frequency (odd-frequency) pair amplitude exists at the
interface without (with) sign change of the pair potential.

In order to understand the angular dependence of the pair amplitude in a
more detail, we define $\hat{f}_{1}$ and $\hat{f}_{2}$ for $-\pi /2<\theta
<3\pi /2$ with $\hat{f}_{1(2)}=f_{1(2)+}(\theta )$ for $-\pi /2<\theta <\pi
/2$ and $\hat{f}_{1(2)}=f_{1(2)-}(\pi -\theta )$ for $\pi /2<\theta <3\pi /2$%
. We decompose $\hat{f}_{1(2)}$ into various angular momentum component as
follows,
\begin{equation}
\displaystyle\hat{f}_{1(2)}=\sum_{m}S_{m}^{(1(2))}\sin [m\theta
]+\sum_{m}C_{m}^{(1(2))}\cos [m\theta ]
\end{equation}%
with $m=2l+1$ for odd-parity case and
$m=2l$ for even parity case with  integer $ l\geq 0$,
%
%
where  $l$ is the quantum number of the angular momentum.
It is straightforward to show that the only nonzero components are
(1) $ C_{2l}^{(2)}$ and $C_{2l+1}^{(1)}$ for $s$ or
$d_{x^{2}-y^{2}}$-wave, (2) $ S_{2l}^{(2)}$ and $S_{2l+1}^{(1)}$ for
$d_{xy}$-wave, (3) $C_{2l+1}^{(2)}$ and $C_{2l}^{(1)}$ for
$p_{x}$-wave, and (4) $S_{2l+1}^{(2)}$ and $ S_{2l}^{(1)}$ for
$p_{y}$-wave junctions, respectively. \par


Below we illustrate the above results by numerical calculations. As
typical examples, we choose $s$-wave and $p_{x}$-wave pair
potentials. Although both $\hat{f}_{1}$ and $\hat{f}_{2}$ have many
components with different angular momenta, we focus on the lowest
values of $l$. We denote $E_{s}(i\omega _{n},x)=C_{0}^{(2)}$,
$E_{px}(i\omega _{n},x)=C_{1}^{(2)}$, $O_{s}(i\omega
_{n},x)=C_{0}^{(1)}$, and $O_{px}(i\omega _{n},x)=C_{1}^{(1)}$ and choose $%
i\omega _{n}=i\pi T$ with temperature $T=0.05T_{C}$. For the
$s$-wave case, the pair potential is suppressed only for high
transparent junctions (see Fig.1a). The odd-frequency component
$O_{px}(i\pi T,x)$ is enhanced for $Z=0$ near the interface where
the pair potential $\Delta (x)$ is suppressed, while the
even-frequency component $E_{s}(i\pi T,x)$ remains almost constant
in this case. For low transparent junctions, the magnitude of
$O_{px}(i\pi T,x)$ is negligible (see Fig.1b). For the $p_{x}$-wave
junction with $Z=0$, although the odd-frequency component
$O_{s}(i\pi T,x)$ is enhanced near the interface, it is smaller than
the even-frequency one $E_{px}(i\pi T,x)$ (see Fig.1c). For the low
transparent junction, the magnitude of $O_{s}(i\pi T,x)$ is strongly
enhanced near the interface and becomes much larger than the
magnitude of $E_{px}(i\pi T,x)$ (see Fig.1d). It is a remarkable
fact that the MARS is reinterpreted as the manifestation of the
odd-frequency pair amplitude at the interface. \par
%
\begin{figure}[tb]
\begin{center}
\scalebox{0.8}{
\includegraphics[width=9cm,clip]{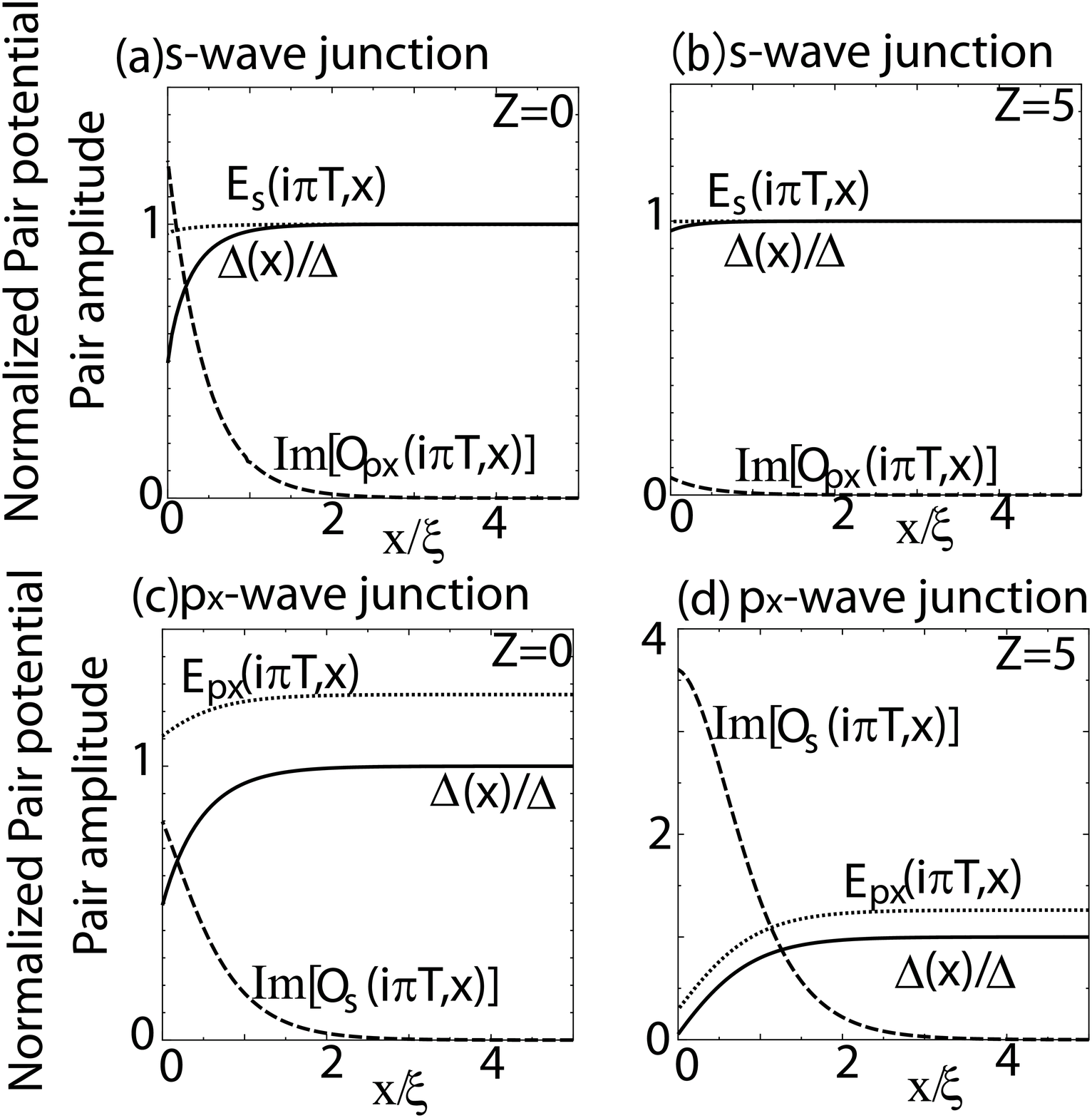}
}
\end{center}
\caption{Spatial dependence of the normalized pair potential (solid
line) even-frequency pair amplitude (dotted line) and odd-frequency
pair amplitude (dashed line). For a bulk state,  we choose
ESE $s$-wave and ETO  $p_{x}$-wave superconductor.
(a) and (c): fully transparent
junctions with $Z=0$. (b) and (d): low transparent junctions with
$Z=5$. $E_{s}(i\protect\pi T,x)$ and $E_{p_{x}}(i\protect \pi T,x)$
are the even-frequency components of the pair amplitude, $O_{s}(i
\protect\pi T,x)$ and $O_{px}(i\protect\pi T,x)$
are the odd-frequency components. The
distance $x$ is normalized by the $v_{F}/\Delta $. } \label{fig:1}
\end{figure}
If we choose $d_{xy}$-wave pair potential,
the magnitude of the odd-frequency component is enhanced at the interface
for large $Z$. In this case, the OSO state is
generated while the OTE state
is generated for $p_{x}$-wave case. The $s$-wave isotropic component that is
robust against the impurity scattering \cite{Sigrist,Golubov2006} appears
only in the $p_{x}$-wave case. The OSO state cannot penetrate into a 
diffusive normal metal while the OTE state can. Thus we can naturally
understand the presence of proximity effect with MARS in $p_{x}$-wave
junctions and its absence in $d_{xy}$-wave junctions
\cite{proximity,Golubov2006}.
\par
In the following, we discuss the manifestations of the
odd-frequency pairing in the Josephson effect
between even- and odd-frequency superconductors.
For this purpose, we
first extend the above discussion regarding Eqs. (1)-(5)
to the case of bulk odd-frequency pairing in a superconductor. In this case,
$\bar{\Delta}_{\pm}(x)$ depends on a Matsubara frequency
$\bar{\Delta}_{\pm}(x)=\bar{\Delta}_{\pm}(x,\omega_{n})$ with
$\bar{\Delta}_{\pm}(x,\omega_{n})=-\bar{\Delta}_{\pm}(x,-\omega_{n})$.
One can show that $f_{2\pm}$ ($f_{1\pm}$) becomes an odd-frequency
(even-frequency) pair amplitude and ETO (ESE) pairing state is
induced at the interface for the OTE (OSO) bulk superconductor.
For the low transparency limit, only  odd-frequency states exist at
the interface for
$\Delta_{\pm}(x,\omega_{n})=\Delta_{\mp}(x,\omega_{n})$ where the
sign change of the pair potential is absent.
On the other hand, in the presence of the sign change, $i.e.$,
$\Delta_{\pm}(x,\omega_{n})=-\Delta_{\mp}(x,\omega_{n})$, only
even-frequency states exist. \par
To summarize the above results, we present
the relation between the order parameter symmetry in the
bulk superconductor and at the interface for $T_{m} \rightarrow 0$ (free surface).
As shown in the Table I below, there are  eight distinct 
cases 
which correspond to different combinations of the bulk pairing symmetry and
the behavior of the orbital part of the bulk pair potential with respect to
reflection from the interface.

\begin{center}
\begin{table}[h]
\begin{tabular}{|c|p{3.5cm}|p{2cm}|p{2cm}|}
\hline
& bulk state & sign change & interface state \\ \hline
(1)& ESE ($s$ or $d_{x^{2}-y^{2}}$-wave) &   No  &  ESE \\ \hline
(2)& ESE ($d_{xy}$-wave) &  Yes &  OSO \\ \hline
(3)& ETO ($p_{y}$-wave) &  No &  ETO \\ \hline
(4)& ETO ($p_{x}$-wave) &  Yes&  OTE \\ \hline
(5)& OSO ($p_{y}$-wave)&  No  &  OSO \\ \hline
(6)& OSO ($p_{x}$-wave)&  Yes &  ESE \\ \hline
(7)& OTE ($s$ or $d_{x^{2}-y^{2}}$-wave) &  No &  OTE \\ \hline
(8)& OTE ($d_{xy}$-wave) & Yes&  ETO \\ \hline
\end{tabular}
\caption{The relation between the symmetry of the
bulk superconductor and that of the pair amplitude
at the interface in the low transparent limit.
The allowed symmetry of the Cooper pair
in accordance with Pauli's rule  are
even-frequency spin-singlet even-parity (ESE),
even-frequency spin-triplet odd-parity (ETO),
odd-frequency spin-singlet odd-parity (OSO),
and odd-frequency spin-triplet even-parity (OTE).
}
\end{table}
\end{center}

Let us discuss the Josephson coupling at the interface between even-frequency
and odd-frequency superconductors to the first order in the interface transparency coefficient $T_{m}$,
assuming that spin-flip scattering at the interface is absent.
According to the Table I, these are 16 possible combinations of pairing symmetries in two superconductors.
Due to the difference of the spin structure of Cooper pairs, the Josephson coupling is absent
for the following combinations: (1)-(7), (1)-(8), (2)-(7), (2)-(8), (3)-(5),
(3)-(6), (4)-(5), and (4)-(6).
The Josephson coupling is also absent for the combinations
(1)-(5), (2)-(6), (3)-(7) and (4)-(8),
since the odd- and even-frequency pairing states are realized on both sides of the interface.
This result is consistent with the previous prediction \cite{Abraham}.
The remaining four combinations (1)-(6), (2)-(5),  (3)-(8), and  (4)-(7)  are 
worthy of remark.
As seen from the above Table, the pairing symmetries on both sides of the interface are the same,
ESE, OSO, ETO and OTE, respectively. As a result, the Josephson current
can flow across the interface in these cases. \par
%
These results can be applied to actual materials. Recently, Fuseya $et.$
$al.$ predicted that the OSO state could be realized in CeCu$_{2}$Si$_{2}$
and CeRhIn$_{5}$ \cite{Fuseya}. It is consistent with some 
experiments \cite{Odd}.
Here, we propose a robust check of pairing symmetry using the Josephson effect 
between ESE (conventional low $T_c$) and OSO (e.g. CeCoIn$_{5}$ \cite{Izawa}) superconductors.
If two ESE superconductors are attached to opposite (parallel) sides of an OSO sample,
the ESE order parameters induced at the two interfaces in OSO will have 
opposite signs.
Then the structure will behave as a $\pi$-junction. Detection of a $\pi$-shift would thus
be an unambiguous signature of OSO pairing symmetry, similar to the phase-sensitive tests 
of d-wave symmetry in high $T_c$ cuprates \cite{Tsuei}.
As regards the OTE state, the promising system is a 
diffusive ferromagnet /spin-singlet $s$-wave (DF/S) hybrid structure,
where OTE state is induced in DF.
Recent calculation of the Josephson effect in spin-triplet $p$-wave / DF/S
junctions \cite{Yokoyama} is consistent with the present prediction.
From this point of view, it is of interest to study junctions 
between Sr$_{2}$RuO$_{4}$\cite{Maeno}
and DF/S hybrids.
\par

In summary, using the quasiclassical Green's function formalism, we
have shown that the odd-frequency pairing state is generated near
normal metal / even-frequency superconductor (N/S) interfaces
in the absence of  spin flip scattering.  When the pair potential in the bulk
has an even-frequency symmetry (spin-singlet even-parity ESE or
spin-triplet odd-parity ETO state), the resulting order parameter at
the interface has an odd-frequency symmetry (spin-singlet odd-parity
OSO or spin-triplet even-parity OTE state), in agreement with the
Pauli principle. On the other hand, if a superconductor has an
odd-frequency (OSO or OTE) order parameter in the bulk, then,
respectively, ESE or ETO pairing state should be induced near the
interface. It follows from the above results that the Josephson
coupling may occur between odd- and even-frequency
superconductors and phase-sensitive tests can be performed to search 
for an odd-frequency
superconducting state. Though we explicitly studied the N/S
junctions only, the odd-frequency pairing state is also expected
near impurities and within Abrikosov vortex cores in even-frequency
superconductors.
%
This implies that the odd-frequency pairing is not at
all a rare situation as was previously considered 
but should be a key concept for understanding the physics of non-uniform
superconducting systems.

One of the authors Y.T. expresses his sincerest gratitude to
discussions with K. Miyake, N. Nagaosa and K. Nagai.
This work is supported by Grant-in-Aid for Scientific Research
(Grant No. 17071007, 17071005  and 17340106)
from the Ministry of Education, Culture, Sports,
Science and Technology of Japan.


\end{document}